# Relational Databases Ingestion into a NoSQL Data Warehouse

Fatma Abdelhedi
CBI² research laboratory, Trimane,
Paris, France.
E-mail : fatma.abdelhedi@trimane.fr

Rym Jemmali
Toulouse Institute of Computer Science
Research (IRIT), CBI²- Trimane,
Paris, France
E-mail : rym.jemmali@trimane.fr

Gilles Zurfluh
IRIT, Capitole University,
Toulouse,France
E-mail : gilles.zurfluh@ut-capitole.fr

*Abstract—* The digital transformation of companies has led to the evolution of databases towards Big Data. Our work is part of this context and concerns more particularly the mechanisms to extract datasets stored in a Data Lake and to store the data in a Data Warehouse. The latter will allow, in a second time, decisional analysis. In this paper, we present the extraction mechanism limited to relational databases. To automate this process, we used the Model Driven Architecture (MDA), which offers a formalized environment for schema transformation. From the physical schemas describing a Data Lake, we propose transformation rules that allow the creation of a Data Warehouse stored on a document-oriented NoSQL system. An experimentation of the transformation process has been performed on a medical application.

*Keyword-Data Lake; Data Warehouse; NoSQL; Big Data; Relational Database; MDA; QVT.*

## I. INTRODUCTION

Due to the considerable increase of data amount generated by human activities, Data Lakes have been created within organizations, often spontaneously, by the physical grouping of datasets related to the same activity. A Data Lake [1] is a massive grouping of data consisting of structured or unstructured datasets. These datasets generally have the following characteristics: (1) they can be stored on heterogeneous systems, (2) each of them is exploited independently of the others, (3) some of them can contain raw data, i.e., data stored in their original form and without being organized according to the use that will be made of them, (4) the types and formats of the data can vary. In practice, a Data Lake can group together different datasets [2] such as relational databases, object databases, Comma Separated Values (CSV) files, texts, spreadsheet folders, etc. The massive data contained in a Data Lake represents an essential reservoir of knowledge for business decision makers. This data can be organized according to a multidimensional data model in order to support certain types of decision processing [3]. For example, in the French health sector, a Data Lake has been created by the French public health insurance company under the name of "Espace Numérique de Santé" (ENS); it includes the electronic health records of insured persons, health questionnaires, and care planners. However, the heterogeneity of storage systems
combined with the diversity of content in the Data Lake is a major obstacle to the use of data for decision-making. To manipulate a Data Lake, a solution consists in ingesting the data into a Data Warehouse and then transforming it (grouping, calculations, etc.). Ingestion is a process that consists in extracting data from various sources and then transferring them to a repository where they can be transformed and analyzed. For example, in [4] massive data from various sources are ingested into a Data Warehouse and exploited in the context of information retrieval on the Web. Other works have introduced the concept of polystore, which preserves the initial data sources (no ingestion) and allows querying them by creating "data islands", each of which contains several systems sharing a common query language. For example, all relational databases are connected to the "relational island", which is queried using standard SQL. This solution, developed in particular in the BiGDAWG [5] and ESTOCADA [6] projects, keeps the data in their native formats.

Our work aims at performing decisional processing on a Data Lake. This problem is part of a medical application in which, massive data are stored in a Data Lake that will be used by medical decision makers. We have chosen to ingest the data from the Data Lake into a Data Warehouse that will later be reorganized for Big Data Analytics. This paper is limited to the ingestion of relational databases and excludes for the moment other forms of datasets present in the Data Lake.

Our paper is organized as follows: in the following Section 2, we present the medical application that justifies our work's purpose. Section 3 describes the context of our study as well as our research problem which aims at facilitating the querying of data contained in a Data Lake by decision makers. Section 4 describes the databases metamodels used in our application. Section 5 presents our contribution which consists in formalizing with the Model Driven Architecture (MDA), the process of transforming the Data Lake databases into a unique NoSQL Data Warehouse. Section 6 describes an experimentation of the proposed process based on our medical application. Section 7 contrasts our proposal with related works. Finally, Section 8 concludes this paper and highlights possible directions for exploring the continuity of the work.

## II. CONTEXT OF WORK

In this section, we present the case study that motivated our work, as well as the problem addressed in this paper.

### A. Case Study

Our work is motivated by a project developed in the health field for a group of private health insurance companies. These insurance companies, stemming from the social and solidarity economy, propose to their customers a coverage of the medical expenses, which comes in complement of those refunded by a public institution: the public health insurance fund.

To ensure the management of their clients, these private health insurance companies are facing a significant increase in the volume of data processed. Indeed, some of these companies carry out all the computer processing related to a record. A digital health platform (ENS) has been developed by the the public authority to store the medical data of each insured person. Private health insurance companies can extract data from the ENS to process the files of their clients and, more broadly, carry out analyses of any kind (in compliance with confidentiality rules). For each insured person, the ENS contains administrative data, medical files (civil status, medical imaging archives, reports, therapeutic follow-ups, etc.), the history of refunds and questionnaires. When the ENS is fully deployed at the national level, its volume will be considerable since it concerns 67 million insured persons.

In the context of this project, the ENS constitutes a real Data Lake because of (1) the diversity of data types and formats (2) the volumes stored which can reach several terabytes and (3) the raw nature of the data. The objective of the project is to study the mechanisms for extracting data from the ENS and organizing it to facilitate analysis (Big Data Analytics).

### B. Problematic

Our work aims to develop a system allowing private health insurance companies to create a Data Warehouse from a Data Lake. This paper deals more specifically with the mechanisms of extraction and unification from commonly used databases, we limit the framework of our study as follows:

-The ENS Data Lake is the source of the data; in this paper, we voluntarily reduce its content to relational databases. Indeed, this category of datasets represents an important part of the ENS data:

- The generated Data Warehouse is managed by a document-oriented NoSQL system. This type of system offers (1) a great flexibility to reorganize objects for analysis and (2) good access performances to large volumes of data (use of MapReduce).

To achieve our goal, each database in the Data Lake is extracted and converted into another model to allow its storage in the Data Warehouse. We do not address here the problems related to the selective extraction of data and their aggregative transformation.

To test our proposals, we have developed a Data Lake with several relational databases managed by MySQL[7] and PostgreSQL[8] systems. These databases contain respectively data describing the follow-up of the insured and the processing of the files in a medical center. The available metadata are limited to those accessible on the storage systems (absence of ontologies for example). The Data Warehouse, which is supported by an OrientDB [9] platform, must allow the analysis of the care pathways of insured persons with chronic pathologies. We chose the OrientDB system to store the Data Warehouse. Indeed, this document-oriented NoSQL system allows to consider several types of semantic links such as association, composition and inheritance links; it is thus well adapted to our case study where the richness of the links between objects constitutes an essential element for decisional processes.

### III. OVERVIEW OF OUR SOLUTION

Although a Data Lake can contain files of any format, we focus in this paper on the extraction of relational databases and the feeding of a NoSQL Data Warehouse. Several works have dealt with the transfer of a relational database to a NoSQL database. Thus, some works have proposed algorithms for converting relational data to document-oriented systems, such as MongoDB [10]; however, these works transform relational links into embedded documents or DBRef links. However, these NoSQL linkage solutions are not satisfactory with respect to object systems[11]. Moreover, to our knowledge, no study has been conducted to convert several relational databases contained in a Data Lake into a NoSQL Data Warehouse.

In our ingestion process, we have defined three modules: the first module named CreateDW, the second ConvertLinks and the last one MergeClasses. We used a Data Lake as a source database for our process, which we limited in this paper to Relational databases and as a target database, we used a NoSQL Data Warehouse that will contain the final processed data. We named our process RDBToNoSDW. Our proposal is based on Model Driven Architecture (MDA) which allows to describe separately the functional specifications and the implementation of an application on a platform. Among the three models present in MDA (CIM, PIM and PSM), we are located at the PSM level where the logical schemas are described. We also use the declarative language Query View Transformation (QVT) [12] specified by the Object Management Group (OMG) [13], which allows us to describe the ingestion of data by model transformations.

To use the MDA transformation mechanism, we proposed two metamodels describing respectively the source and target databases. From these metamodels, we specified the transformation rules in QVT language to ensure data ingestion.

### IV. METAMODELING

We present successively our metamodels proposal of a source Relational database and a target document-oriented NoSQL database.

### A. Relational Metamodel

The Data Lake, source of our process, can contain several relational databases. A relational database contains a set of tables made of a schema and an extension. The schema of a table contains a sequence of attributes. The extension is composed of a set of rows grouping attribute values. Among the attributes of a table, we distinguish the primary key whose values identify the rows and the foreign keys, which materialize the links. Figure 1 represents the Ecore[14] metamodel of a relational database.

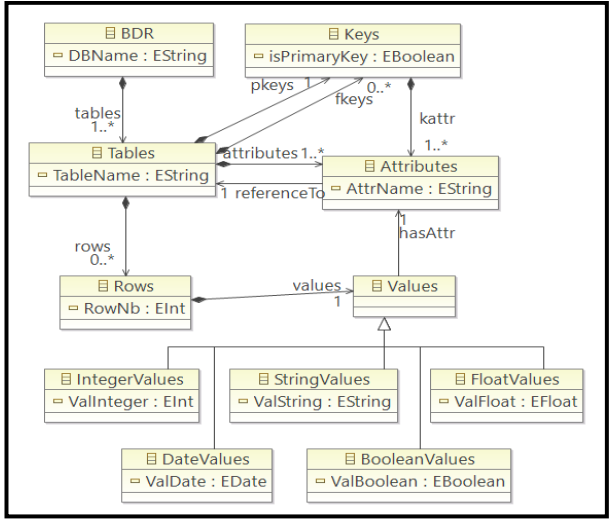

Figure 1. Metamodel of a relational database

### B. Document-Oriented NoSQL Metamodel

The target of our process corresponds to the Data Warehouse represented by a NoSQL database. A document-oriented NoSQL database contains a set of classes. Each class gathers objects that are identified (by a reference) and composed of couples (attribute, value); a value is defined by a type, it can be either multivalued or structured. We distinguish a particular type, the reference, whose values make it possible to link the objects. These concepts are represented in Figure 2 according to the Ecore formalism.

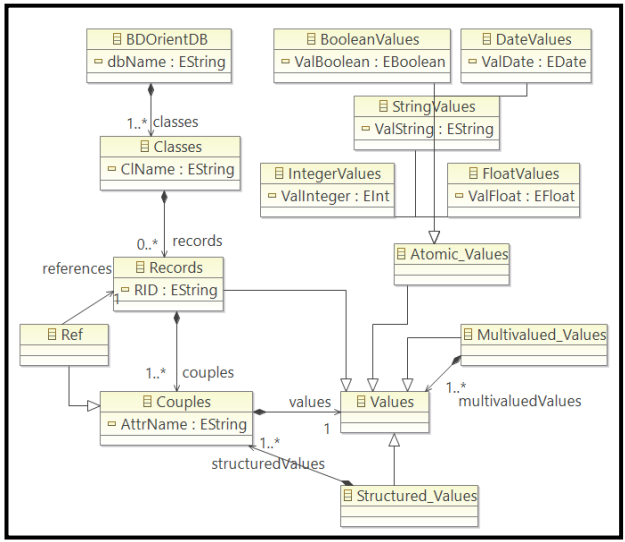

Figure 2. Metamodel of a document-oriented NoSQL database

## V. DATA MANAGEMENT

This involves transferring relational databases from the Data Lake to a NoSQL database corresponding to the Data Warehouse. To carry out this ingestion process, we have defined three modules that will successively ensure (1) the transformation of relational data into NoSQL data (CreateDW module), (2) the conversion of relational links (foreign keys) into references (ConvertLinks module) and (3) the merging of tables containing objects of the same semantics (MergeClasses module).

### A. CreateDW Module

This module transforms each relational database of the Data Lake into a unique NoSQL database according to the MDA approach. The NoSQL warehouse being unique, it will contain the data coming from the different relational databases of the Data Lake. The application of a set of transformation rules defined on the metamodels of Section 4, generates a set of classes in a NoSQL database. We informally present the rules that have been expressed in the QVT language.

**Rule 1:** Each table in a relational database is transformed into a class in the NoSQL database. To avoid synonymy, the name of the class will be prefixed by the name of the original database.

**Rule 2:** Each row of a table, associated with its schema, is transformed into a record in the corresponding target class; the record then contains a set of couples (attribute, value). The primary key is stored as any attribute. At this stage, the foreign keys are also stored with their relational values; they will be converted into references by the ConvertLinks module. These two rules, that we formalized in QVT language, are applied for each relational database of the Data Lake and feed the NoSQL DB; we will present their syntax in Figure 4 of the experimentation section. In parallel with the application of these transformation rules, an algorithmic processing allows to record metadata; these metadata match each relational primary key with the Record Identifier (RID) of the corresponding record in the NoSQL database.

### B. ConvertLinks Module

In the standard object-oriented systems[15], links are materialized by references. Since this principle is used in NoSQL systems, it is necessary to convert relational foreign keys that have been transferred to the Data Warehouse into references.

The mechanism we have developed in ConvertLinks is not based on the expression of MDA rules but corresponds to an algorithmic process. In the NoSQL database, all records of a class are systematically marked with identifiers (RID for Record ID). During the transfer of data into the records, the relational primary and foreign keys were transferred in the form of pairs (attribute, value). Thus, thanks to the metadata recorded by the previous CreateDW module, the values of the foreign keys are converted into RID.

## C. MergeClasses Module

The ingestion of data from the Data Lake has been done by transferring the data from the different relational databases into the NoSQL database. However, it is common for tables with the same semantics to be transferred from different relational databases; these tables are said to be "equivalent", for example the DB1-Insured table and the DB2-Patients table containing data on the insured. It is therefore useful to group the data contained in "equivalent" tables within a single class of the NoSQL database. To achieve this grouping, we relied on an ontology establishing the correspondences between the terms of the relational databases contained in the Data Lake. This ontology is provided by relational data administrators bringing their business expertise. These administrators, after consultation, have associated the tables considered as semantically equivalent.

Using this ontology, the MergeClasses module creates new classes in the NoSQL database; each of these classes groups the data from the various equivalent tables. This process is not limited to a union operation between records.

In fact, distinct records concerning the same entity can have complementary attributes that will be combined in a single record.

## VI. IMPLEMENTATION AND TECHNICAL ENVIRONMENT

In this section, we describe the techniques used to implement the RDBToNoSDW process. We used the Eclipse Modeling Framework (EMF) technical environment that is suitable for modeling, metamodeling, and transforming models. EMF has the Ecore metamodeling language to create and manipulate metamodels. Ecore is based on XMI to instantiate models and QVT to transform metamodels. Algorithmic processing was coded in Java because of its compatibility with Eclipe which is the development environment used.

The CreateDW module of our process generates a unique NoSQL database from the Data Lake databses. It uses a relational metamodel and a NoSQL metamodel as represented with Ecore in Figures 2 and 3. The instantiation of the two relational databases is done with the XMI language. Figure 3 shows the XMI instantiation of a source relational database. The transformation rules have been translated with the QVT language (Figure 4) and apply to all relational databases, independently of the RDBMS used.

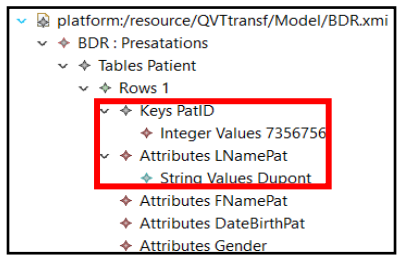

Figure 3. XMI instantiation of a source relational database

```
//Transform a relational database into an OrientDB database
mapping BDR::RDBtoODB(): BDOrientDB{
dbName:=self.DBName;
classes:= self.tables.map toTable();
}

//Transform a relational table into an OrientDB class
mapping RDB::Tables:: toTable():ODB::Classes{
ClName := self.TableName;
records:= self.rows.map toDoc();
}

//Transform a row into an OrientDB Record
mapping RDB::Rows:: toDoc():ODB::Records{
couples:=self.hasAttributes.map toCouple();
```

Figure 4. QVT transformation rules from relational to NoSQL databases

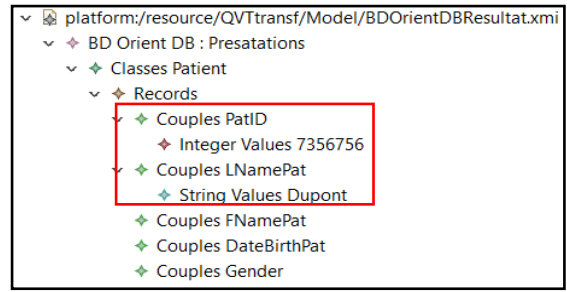

Figure 5. Result XMI file of a target NoSQL database (after QVT rules execution)

The result of applying the QVT transformation rules (Figure 4) is shown in Figure 5.

At the end of the execution of the CreateDW module, we obtain a NoSQL Data Warehouse containing a set of classes as shown in Figure 6. Each of them corresponds to a relational table without any filtering having been carried out (possible presence of "equivalent" tables stored in different databases from the source).

Figure 6. Extract from the list of the Data Warehouse classes stored in OrientDB

The ConvertLinks module converts relational foreign keys into RID. For example, we consider a table "Patient" containing a patient information's with a field "Doctor" representing a foreign key. This field will, therefore, be converted to a reference (RID). Figure 7 shows a record of the "Patient" class after running the ConvertLinks module.

```
{
  "@type": "d",
  "@rid": "#26:0",
  "@version": 1,
  "@class": "Analysis_Patients",
  "Email": "ramon.saadi@gmail.com",
  "FNamePat": "Ramon",
  "LNamePat": "Saadi",
  "NoPat": "45657709",
  "Doctor": "#22:0"
}
```

Figure 7. Extract from the "Patient" class after limks convertion

Finally, the MergeClasses module groups the records of the classes considered as "equivalent" based on the ontology provided by the experts. Figures 8 and 9 represent respectively two records from two classes "ServiceProvision_Insured" and "Analysis_Patients". The two records, having in common several semantically equivalent attributes, will be merged into a single record stored in the same class "Insured_DW" as shown in Figure 10.

```
{
  "@type": "d",
  "@rid": "#34:0",
  "@version": 1,
  "@class": "ServiceProvision_Insured",
  "Gender": "M",
  "FNameIns": "Ramon",
  "LNameIns": "Saadi",
  "NoInsured": "45657709",
  "Spouse": "#36:0"
}
```

Figure 8. Record fron the « ServiceProvision_Insured » class

```
{
  "@type": "d",
  "@rid": "#26:0",
  "@version": 1,
  "@class": "Analysis_Patients",
  "Email": "ramon.saadi@gmail.com",
  "FNamePat": "Ramon",
  "LNamePat": "Saadi",
  "NoPat": "45657709",
  "Doctor": "#22:0"
}
```

Figure 9. Record fron the « Analysis_Patients » class

```
{
  "@type": "d",
  "@rid": "#62:0",
  "@version": 1,
  "@class": "Insured_DW",
  "Email": "ramon.saadi@gmail.com",
  "FNameIns": "Ramon",
  "LNameIns": "Saadi",
  "NoInsured": "45657709",
  "Doctor": "#22:0",
  "Gender": "M",
  "Spouse": "#36:0"
}
```

Figure 10. Record fron the new created « Insured_DW » class

Figure 11. Extract from the "Insured_DW" class

Figure 11 represents an extract of the new class "Insured_DW" containing a record resulting from merging records belonging to the two classes "ServiceProvision_Insured" and "Analysis_Patients".

## VII. RELATED WORKS

In this section, we present research work on extracting data from a Data Lake and more specifically data from several relational databases and creating a NoSQL Data Warehouse. The advent of Big Data has created several challenges for the management of massive data; among these we find the creation of architectures to ingest massive data sources as well as the integration and transformation of these massive data (Big Data) allowing their subsequent query. In this sense, some works have focused on the proposal of architectures (physical and logical) allowing the use and the management of Data Lakes. The work in [16] proposes an approach to structure the data of a Data Lake by linking the data sources in the form of a graph composed of keywords. Other works propose to extract the data of a Data Lake from the metamodels of the sources. The authors in [17] have proposed a metamodel unifying NoSQL and relational databases. There are several formalisms [18] to express model transformations such as the QVT standard, the ATL language [19], which is a non-standardized model transformation language more or less inspired by the QVT standard of the Object Management Group, etc.

Other works have studied only the transformation of a relational database into a NoSQL database. Thus in [20, 21] the authors developed a method to transfer data from relational databases to MongoDB. This approach translates the links between tables only by nesting documents. In [22], the authors present MigDB, an application that converts a relational database (MySQL) to a NoSQL one (MongoDB. This conversion is done over several steps: transforming tables into JSON files, then transmitting each JSON file to a neural network. This network allows to process the links at the JSON file level, either by nesting or by referencing. This work considers association links only. The same is true in [23], where the authors propose a method for transferring relational databases to MongoDB by converting the tables into CSV files that are then imported into MongoDB. However, the proposed method simply converts tables into MongoDB collections without supporting the various links between tables.

Our solution is based on the metamodeling of the sources of a Data Lake, the transformation of these metamodels thanks to the QVT standard and then the creation of a NoSQL Data Warehouse stored under OrientDB allowing to query the data of the Data Lake.

## VIII. CONCLUSION

We have proposed a process to ingest data from a Data Lake into a Data Warehouse. The Data Lake contains several databases. This paper focuses on a specific problem, we have limited the content of the Data Lake to relational databases.

Three modules ensure the ingestion of the data. The CreateDW module transforms each relational database into a unique NoSQL database by applying MDA rules. This mechanism will be extended to transform other types of databases stored in the Data Lake. The ConvertLinks module translates relational links (keys) into references in accordance with the principles of object databases supported by the OrientDB system. Finally, the MergeClasses module merges semantically equivalent classes from different Data Lake databases; this merge is based on an ontology provided by business experts.

Currently, we are continuing our work on the ingestion of data from a Data Lake by extending it to other types of data sources, ingesting and processing data coming from CSV files, NoSQL databases (document and column-oriented databases) and text files. Indeed, these types of files are present in the Data Lake of our medical case study.